\documentclass[aps,notitlepage,superscriptaddress,amsmath,amssymb,longbibliography]{revtex4-1}
\usepackage{xcolor}
\usepackage{graphicx}
\usepackage{soul}
\begin{document}

% Use the \preprint command to place your local institutional report
% number in the upper righthand corner of the title page in preprint mode.
% Multiple \preprint commands are allowed.
% Use the 'preprintnumbers' class option to override journal defaults
% to display numbers if necessary
%\preprint{}

%Title of paper
\title{Oblique droplet impact onto a deep liquid pool}

% repeat the \author .. \affiliation  etc. as needed
% \email, \thanks, \homepage, \altaffiliation all apply to the current
% author. Explanatory text should go in the []'s, actual e-mail
% address or url should go in the {}'s for \email and \homepage.
% Please use the appropriate macro foreach each type of information

% \affiliation command applies to all authors since the last
% \affiliation command. The \affiliation command should follow the
% other information
% \affiliation can be followed by \email, \homepage, \thanks as well.
%\author{}
%\email[]{Your e-mail address}
%\homepage[]{Your web page}
%\thanks{}
%\altaffiliation{}
%\affiliation{}

\author{Sten A. Reijers \email[]{s.a.reijers@utwente.nl}}
\affiliation{Physics of Fluids Group, Faculty of Science and Technology, MESA+Institute, University of Twente, P.O. Box 217, 7500 AE Enschede, The Netherlands}
\author{Bo Liu}
\affiliation{Advanced Research Center for Nanolithography (ARCNL), Science Park 110, 1098 XG Amsterdam, The Netherlands}
\author{Detlef Lohse}
\affiliation{Physics of Fluids Group, Faculty of Science and Technology, MESA+Institute, University of Twente, P.O. Box 217, 7500 AE Enschede, The Netherlands}
\author{Hanneke Gelderblom \email[]{h.gelderblom@tue.nl}}
\affiliation{Physics of Fluids Group, Faculty of Science and Technology, MESA+Institute, University of Twente, P.O. Box 217, 7500 AE Enschede, The Netherlands}
\affiliation{Department of Applied Physics, Eindhoven University of Technology, Den Dolech 2, 5600 MB, Eindhoven, Netherlands}

%Collaboration name if desired (requires use of superscriptaddress
%option in \documentclass). \noaffiliation is required (may also be
%used with the \author command).
%\collaboration can be followed by \email, \homepage, \thanks as well.
%\collaboration{}
%\noaffiliation

\date{\today}

\begin{abstract}
The oblique impact of a liquid droplet onto a deep liquid pool is studied numerically with the adaptive volume-of-fluid solver Basilisk. The splashing threshold, cavity formation, cavity evolution and the maximum cavity dimensions are quantified as a function of the Weber number and the impact angle. We compare the numerical results with recent experimental work by Gielen \textit{et al}. [Phys. Rev. Fluids \textbf{2}, 083602 (2017)]. Similarly to the experimental results, three different impact regimes are observed: deposition of the droplet onto the pool, single-sided splashing in the direction of the impact and splashing in all directions. We show good qualitative and quantitative agreement of the splashing behaviour and cavity formation between the simulations and the experiments. Furthermore, the simulations provide a three dimensional view of the impact phenomenon, give access to velocity and pressure fields, and allow to explore impact parameters that are hard to achieve experimentally.
\end{abstract}

% insert suggested keywords - APS authors don't need to do this
%\keywords{}

%\maketitle must follow title, authors, abstract, and keywords
\maketitle

% body of paper here - Use proper section commands
% References should be done using the \cite, \ref, and \label commands

\section{\label{sec:introduction}Introduction}
The impact of a liquid droplet onto a deep liquid pool induces a broad range of fascinating physical phenomena \citep{rein_1996}. Droplets impacting with a low velocity coalesce with the pool directly or undergoes a coalescence cascade \citep{CHARLES1960105,doi:10.1063/1.870380,leng_2001,leneweit_etal_2005,ZHAO20111109}. Higher impact velocities induce the ejection of a sheet above the pool surface \citep{weiss_yarin_1999,thoroddsen_2002,agbaglah_thoraval_thoroddsen_zhang_fezzaa_deegan_2015}, which upon breakup could result in a splash \citep{doi:10.1063/1.1572815,ray_biswas_sharma_2015,castrejon_JFM_2016,PhysRevFluids.2.083602,doi:10.1039/C7SM01468F}. During impact a cavity develops below the pool surface \citep{doi:10.1063/1.1708605,doi:10.1063/1.1709044,RODRIGUEZ1988121}. The collapse of this cavity \citep{duclaux_caille_duez_ybert_bocquet_clanet_2007,bergmann_vander_meer_gekle_van_der_bos_lohse_2009,PhysRevFluids.2.023601} in combination with capillary wave dynamics can result in bubble entrapment \citep{oguz_prosperetti_1990,doi:10.1063/1.4746793,tran_demaleprade_sun_lohse_2013}, the formation of a fast microjet \citep{longuet-higgins_oguz_1995,PhysRevLett.102.034502} and a Worthington jet \citep{PhysRevE.92.053022}.

The vast majority of droplet impact studies in the literature have considered perpendicular impacts. However, in many practical situations droplets do not impact perpendicularly but under an angle, i.e. obliquely. Examples range from rain impact in nature \citep{doi:10.1121/1.397353,doi:10.1175/JPO-D-17-0172.1} to impact on turbine blades \citep{strathprints36575}, tin catching devices in extreme ultraviolet lithography machines \citep{10.1117/1.JMM.11.2.021109} and spray cooling \citep{KIM2007753} in industry. Precise control of the impact behaviour is key to optimizing these industrial applications in order to prevent damage or contamination.

Recently, the regimes of oblique droplet impact onto deep liquid pools \citep{PhysRevFluids.2.083602,Okawa2008,leneweit2005}, moving films \citep{PhysRevE.92.053005,Alghoul2011} and inclined surfaces \citep{Antonini2014,SIKALO2005661} have been studied experimentally. Furthermore, over the past decades there has been extensive research into analytic modeling of oblique high-velocity impacts onto a inviscid liquid pool \citep{doi:10.1146/annurev.fl.20.010188.001111}. Although these analytic methods allow for a thorough understanding of the early time evolution of e.g. the ejecta sheet \citep{Miloh1991, howison_ockendon_wilson_1991, moore_howison_ockendon_oliver_2012, Moore2013}, numerical studies are required to capture the full impact dynamics.

Most numerical studies so far focused on droplet impact on thin liquid films \citep{RIEBER1999455,GUO201426,XIE2017303,doi:10.1063/1.4996588}, wet walls \citep{NIKOLOPOULOS2007322,MING2014307,CHENG201511}, perpendicular impact onto deep liquid pools \citep{doi:10.1063/1.870332,ray_biswas_sharma_2015,FONTES201818,PhysRevLett.122.014501} or bubble entrainment by impacting droplets \citep{oguz_prosperetti_1990,ray_biswas_sharma_2015,doi:10.1063/1.4992124}. \citet{ray_biswas_sharma_2012} numerically studied the coalescence and splashing regimes for droplets impacting a deep and shallow liquid pool under an impingement angle. As these authors performed two-dimensional simulations, they were only able to capture basic impact features.

\begin{figure}
\centering
\includegraphics[width=0.7\textwidth]{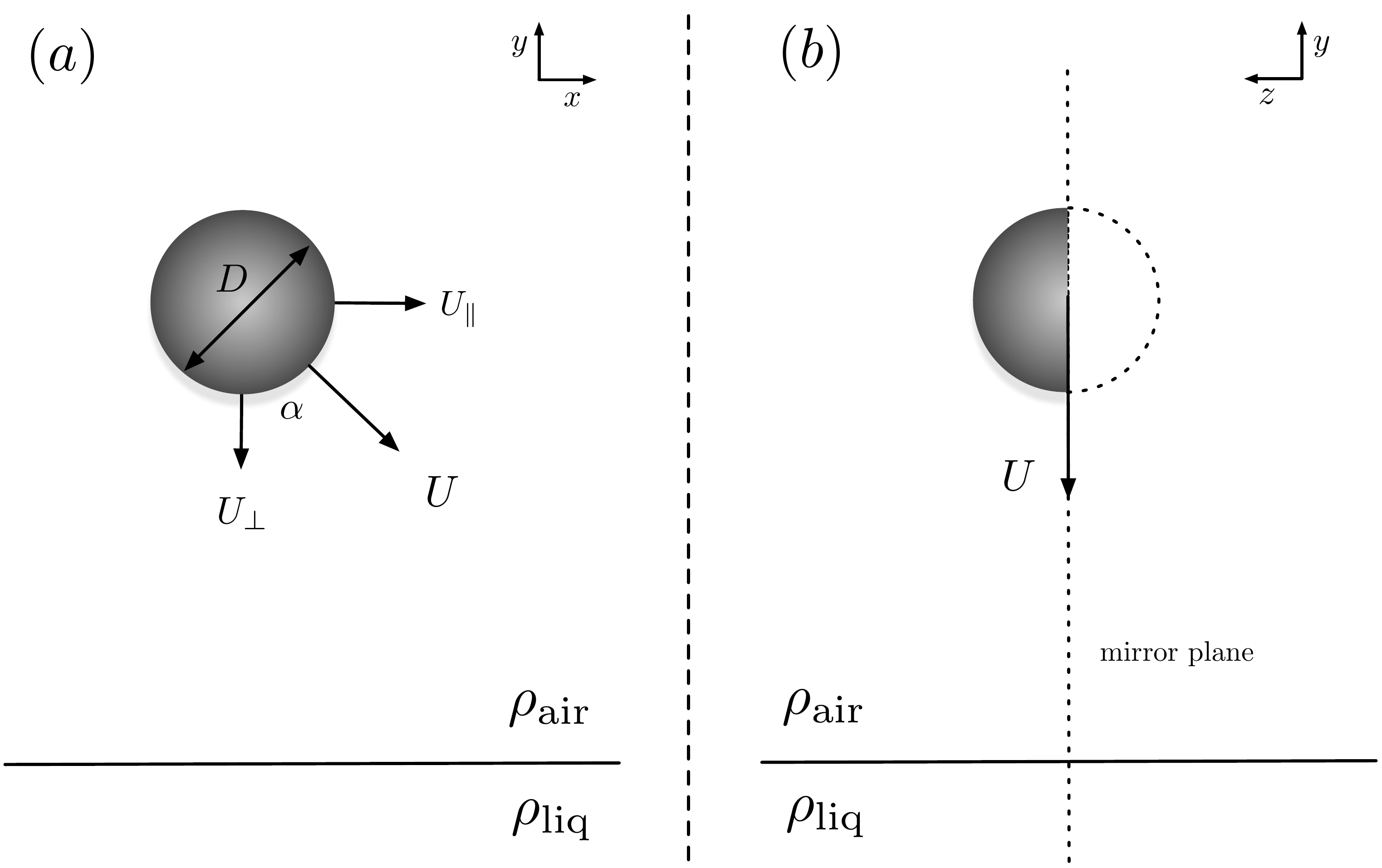}
\caption{\label{fig:sketch} Schematic views of the numerical setup. A liquid droplet with diameter $D$ and density $\rho_{\text{liq}}$ moves with a velocity $U$ under an angle $\alpha$ through air with density $\rho_{\text{air}}$ towards a liquid pool. We define the trailing side as the left and leading side as the right hand side of the droplet. (a) Side-view illustrating the parallel $U_\parallel$ and perpendicular $U_\perp$ velocity components. (b) Front-view with a symmetry plane at $z=0$. By simulating only one half of the domain and using a mirror boundary condition on the plane we reduce the simulation time by approximately a factor of two. }
\end{figure}    

In this work we present a three-dimensional (3D) numerical study of oblique droplet impact onto a deep liquid pool using the open-source volume-of-fluid solver Basilisk \citep{Basilisk2014}. The numerical results are compared to recent experimental results by \citet{PhysRevFluids.2.083602} and show that the method is able to capture the impact dynamics with excellent accuracy. Details of the numerical setup are provided in section \ref{sec:numericalmethod}. In section \ref{sec:results} the simulation results are discussed and compared with the experimental results. In section \ref{sec:conclusion} we discuss and conclude this work.

\section{\label{sec:numericalmethod}Numerical setup}
To solve the 3D incompressible Navier-Stokes equations together with the Volume of Fluid (VOF) method to support two-phase flow, we employ the parallelized octree-adaptive numerical method Basilisk \citep{Basilisk2014}. Basilisk is the successor of the Gerris flow solver and has been extensively validated on problems regarding two-phase complex flows \citep{Popinet2009,PhysRevLett.117.184502, Basilisk2014Tests}. The momentum and VOF equations solved are
\begin{align}
&\boldsymbol{\nabla}\cdot\boldsymbol{u} = 0, \label{eq:incompressiblenavierstokes}\\
&\frac{\partial \boldsymbol{u}}{\partial t} +( \boldsymbol{u}\cdot\boldsymbol{\nabla})\boldsymbol{u} = \frac{1}{\rho}\bigg[-\boldsymbol{\nabla}\textrm{p}+ \boldsymbol{\nabla}\cdot\mu(\boldsymbol{\nabla}\boldsymbol{u}+\boldsymbol{\nabla}\boldsymbol{u}^T) + \rho \boldsymbol{a} + \sigma\kappa\delta_s\boldsymbol{n}\bigg], \label{eq:incompressiblenavierstokesmomentum}\\
&\frac{\partial  f}{\partial t} + \boldsymbol{\nabla}\cdot(f \boldsymbol{u}) = 0, \label{eq:vofadvection}
\end{align}
where $\boldsymbol{u}$ is the velocity field, $\rho$ the density, $\textrm{p}$ the pressure, $\mu$ the viscosity, $\boldsymbol{a}$ optional body forces, $\sigma$ the surface tension, $\kappa$ the surface curvature, $\delta_s$ a delta function on the interface between two fluids, $\boldsymbol{n}$ the surface normal vector on this interface and $f$ the volume fraction field used in the VOF method.

\begin{figure*}
\centering
\includegraphics[width=0.99\textwidth]{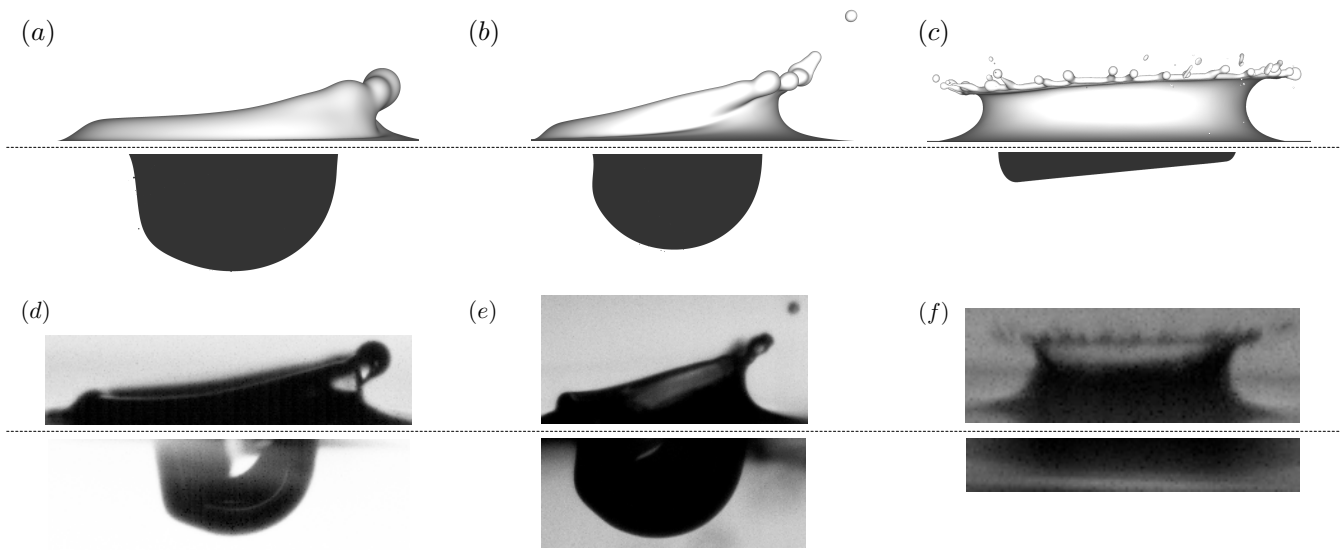}
\caption{\label{fig:simexp} A side-view of three different types of impact behaviour observed above and below the pool surface in the simulations (top panel) and in the experiments (bottom panel) at identical times. From left to right: deposition (first column), single-sided splashing (second column) and omni-directional splashing (last column). Simulation parameters (a) $\text{We} = 187$, $\alpha = 26^\circ$, (b) $\text{We} = 400$, $\alpha = 30^\circ$ and (c) $\text{We} = 600$, $\alpha = 10^\circ$. Experimental parameters (d) $\text{We} = 180.5$, $\alpha = 28.1^\circ$, (e) $\text{We} = 419$, $\alpha = 27.9^\circ$ and (f) $\text{We} = 662$, $\alpha = 0.6^\circ$. Experiments by \citet{PhysRevFluids.2.083602}.}
\end{figure*} 
% Timelapse diagram
\begin{figure*}
\centering
\includegraphics[width=0.99\textwidth]{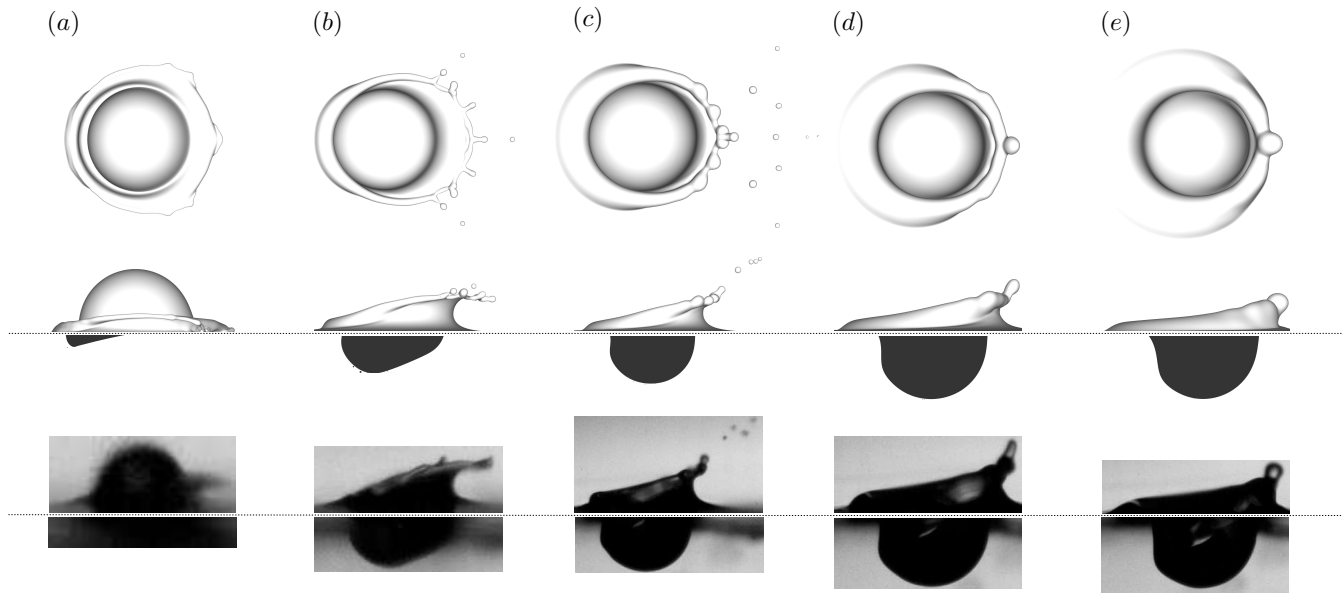}
\caption{\label{fig:timelapse} 
Time series of a single-sided crown splash observed above and below the pool surface in the simulation (two top panels) and in the experiment (bottom panel). Times are non-dimensionalized by $t_i=D/U$, where $t/t_i = 0$ corresponds to the moment of first contact with the pool. Simulation parameters $\text{We} = 400$, $\alpha = 30^\circ$, experiment $\text{We} = 416.5$, $\alpha = 28.5$. (a)  $t/t_i=0.46$: when the droplet hits the surface a sheet ejects on the leading side of the impact area. (b) $t/t_i=2.33$: the sheet evolves into a crown leading to several satellite droplets in the simulation that are hard to detect in the experiment. (c) $t/t_i=8.22$: the crown further expands outwards and the satellite droplets are now clearly visible in both simulations and experiments. (d) $t/t_i=12.15$: the crown retracts due to capillary forces and a finger appears at the leading side. (e) $t/t_i=18.00$: the liquid finger collapses onto the pool due to surface tension forces. Experiments by \citet{PhysRevFluids.2.083602}.}
\end{figure*} 
The fluid equations (\ref{eq:incompressiblenavierstokes},\ref{eq:incompressiblenavierstokesmomentum}) are solved by a finite volume method over an adaptive Eulerian grid with the Bell-Colella-Glaz second-order accurate advection scheme \citep{POPINET2003572,BELL1989257}. The volume fraction field (\ref{eq:vofadvection}) is solved by a piecewise-linear
geometrical scheme by \citet{annurev.fluid.31.1.567} together with a continuum surface force model for the surface tension in (\ref{eq:incompressiblenavierstokesmomentum}) \citep{1992BrackBill,Popinet2009}.

A sketch of the numerical setup is given in figure \ref{fig:sketch}. A spherical liquid droplet with diameter $D$ and density $\rho_{\text{liq}}$ falls with a velocity $U$ under an angle $\alpha$ through air with density $\rho_{\text{air}}$ towards a liquid pool with density $\rho_{\text{liq}}$. We define the leading side as the right- and trailing side as the left-hand side of the droplet. The pool depth is $8D$, the cubic simulation domain edges are $12D$ and the initial distance from the droplet to the pool is $0.1D$. The density and viscosity ratio between the liquid and gas phase in the simulation are $\rho_{\text{liq}}/\rho_{\text{air}} = 1000$, $\mu_{\text{liq}}/\mu_{\text{air}}= 100$, respectively. To ensure stability of the simulation the Courant-Friedrichs-Lewy (CFL) condition is set to $\text{CFL}=0.05$ \citep{Basilisk2014}. Figure \ref{fig:sketch} shows a side view (a) and front view (b) of the simulation. A symmetry plane is enforced in the $y - z$ plane (see figure \ref{fig:sketch}b) allowing us to simulate only half the droplet and save simulation time. Note that this constraint may restrict the development of a Rayleigh-Taylor instability at the rim of the ejecta sheet, therefore if one is interested in a detailed analysis of the fingering instability this constraint should be released. At the bottom of the pool no-slip and impermeability conditions are used, while at the top of the domain free-slip and free-outflow conditions are imposed. For the remaining side planes symmetry boundary conditions are used. The impact location is in the center of the domain, such that the domain walls are of little influence on the early-time dynamics of the impact.

The grid is refined adaptively using the wavelet adaptation method, which is a build-in refinement strategy in the Basilisk framework \citep{annurev-fluid-121108-145637,vanHooft2018}. The maximum refinement level on the grid is bounded to the computational resources and time available, which leads to a maximum refinement level $r_{\text{max}} = 12$ and hence a theoretical maximum grid resolution of $4096^3$ or about $10^5$ cells per droplet cross section when all volumes are refined. The refinement algorithm is invoked every timestep and refines when the wavelet estimated error exceeds $\boldsymbol{u}_{\text{err}} = 10^{-2}$ for the velocity field and $f_{\text{err}} = 5*10^{-3}$ for the fraction field. Initially we refine the interface of the droplet to $r_{\text{max}}$ while the rest of domain stays at $r_{\text{initial}}=6$. With these settings a typical simulation of a single impact event from the moment of impact until the cavity closes takes about a month using a modern computer cluster in parallel. During the simulation we set fraction field values lower than $10^{-4}$ to 0 and values higher than $(1-10^{-4})$ to 1. This smoothing reduces noise caused by tiny droplets and bubbles in the simulation that do not affect the overall dynamics but are costly to track. This filtering may alter the dynamics of bubble entrapment, which is not the focus of the present work.

By neglecting the effect of the ambient air, oblique impact of a droplet onto a liquid pool can be characterized by four dimensionless parameters: the Weber number $\text{We} = \frac{\rho_{\text{liq}} D U^2}{\sigma}$, the Reynolds number $\text{Re}=\frac{\rho_{\text{liq}}UD}{\mu_{\text{liq}}}$, the Froude number $\text{Fr}^2=\frac{U^2}{g D}$ and the impact angle $\alpha$, where $g$ is the gravitational acceleration \citep{PhysRevFluids.2.083602}. We are interested in an impact regime where the influence of Froude and Reynolds is negligible (in the experiments used for comparison \cite{PhysRevFluids.2.083602} $\text{Fr} \in [10^2-10^3]$ and $\text{Re} \in [600-2500]$). Hence in the simulations, the gravitational contribution is neglected, the impact velocity is set to $U=1$ and the liquid viscosity is set such that the Reynolds number is always equal to $\text{Re}=1000$. We then explore a simple two-dimensional phase space $(\text{We}, \alpha)$. Table \ref{tab:simulations} gives a list of the oblique droplet impact simulations performed. We have chosen the simulations parameters $(\text{We},\alpha)$ such that we cover a comprehensive set of splashing phenomena in the phase space. These simulated parameters may therefore not always be a one-to-one match with the experiments performed by \citep{PhysRevFluids.2.083602}. In the analysis of the results it will turn out practical to discriminate between the parallel Weber number $\text{We}_{\parallel} = \frac{\rho_{\text{liq}} D U^2_{\parallel}}{\sigma}$ and the perpendicular Weber number $\text{We}_{\perp} = \frac{\rho_{\text{liq}} D U^2_{\perp}}{\sigma}$, where $U_{\parallel} = U\sin(\alpha)$ and $U_{\perp} = U\cos(\alpha)$ are the velocity components parallel and perpendicular to the pool surface, respectively.
\begin{table}[b]
\centering
\begin{tabular}{|cc|cc|}
\hline
$\text{We}$ & $\alpha (^\circ)$ & $\text{We}$ & $\alpha (^\circ)$ \\
\hline
187.5 & 28& 400 & 60\\
200 & 15& 400 & 75\\
200 & 40& 600 & 10\\
200 & 45& 600 & 20\\
250 & 35& 600 & 45\\
250 & 60& 600 & 75\\
300 & 20& 674 & 28\\
400 & 12& 800 & 20\\
400 & 20& 1000 & 55\\
400 & 30& 1400 & 45\\
400 & 40& 1400 & 60 \\
\hline
\end{tabular}
\caption{\label{tab:simulations}A list of simulations performed with the parameters as described in section \ref{sec:numericalmethod}. The Reynolds number is fixed to $\text{Re} = 1000$ in all simulations and the gravity is disabled.}
\end{table}

\section{\label{sec:results}Results}
In this section we present our numerical results and we show a qualitative and quantitative comparison to the experiments performed by \citet{PhysRevFluids.2.083602}. In \ref{sec:overview} the different types of impact phenomena are discussed and classified. An overview of the early impact dynamics is given in section \ref{sec:dynamics}. The splashing threshold of the crown above the pool is systematically studied in section \ref{sec:splashingregimes}. We discuss the cavity formation and evolution as function of time and compare this result to a theoretical model in section \ref{sec:cavityevolution}. Finally, we show the maximum cavity dimensions below the surface, i.e. the maximal cavity depth, maximal cavity displacement and collapse angle as function of Weber number and impact angle in section \ref{sec:cavitydimensions}. 
\subsection{\label{sec:overview}Impact phenomena}
When a droplet obliquely impacts onto a pool, a cavity forms below the pool surface accompanied by an asymmetric crown above the pool surface. In our simulations we observe three different impact phenomena similar to what has been observed in experiments: deposition, single-sided splashing and omni-directional splashing \citep{PhysRevFluids.2.083602}. Figure \ref{fig:simexp} shows these three phenomena in both the simulation (top panel) and the experiments (bottom panel).

The first column (figure \ref{fig:simexp}a and \ref{fig:simexp}d) shows a deposition event for $\text{We}=187$ and $\alpha = 26^\circ$ (simulation) and $\text{We}=180.5$ and $\alpha=28.1^\circ$ (experiment). Directly after impact a thin sheet ejects on the leading side of the impact cavity. The impact energy of the droplet is not sufficient for this ejecta sheet to develop into a crown on all sides. Instead the ejecta sheet retracts back towards the pool due to capillary forces without any breakup. We observe excellent qualitative agreement of the simulation with the experiment.

The second column (figure \ref{fig:simexp}b and \ref{fig:simexp}e) shows a single-sided splash for $\text{We}= 400$ and $\alpha=30^\circ$ (simulation) and $\text{We} = 419$ and $\alpha=27.9^\circ$ (experiment). Here, the impact energy is large enough for the ejecta sheet to develop several fingers on the leading side of the cavity, which eventually break up into satellite droplets in both the simulation and experiment. The resulting single-sided crown in the simulation compares very well with the overall crown shape of the corresponding experiment. 

Finally in the last column (figure \ref{fig:simexp}c and \ref{fig:simexp}f) an omni-directional splash is observed for $\text{We}= 600$ and $\alpha=10^\circ$ (simulation) and $\text{We} = 662$ and $\alpha=0.6^\circ$ (experiment). In this case the impact energy is large enough to induce a crown with satellite droplets all around the cavity in both the simulation and the experiment.

Figure \ref{fig:timelapse} shows the comparison of a detailed time series of a single-sided splash impact event in the simulation ($\text{We}= 400$ and $\alpha=30^\circ$, two top panels) and in the experiment ($\text{We}= 416.5$ and $\alpha=28.5^\circ$, bottom panel). In addition, the simulation provides a top view of the impact event. Time increases from left to right, with $t/t_i=0$ being the moment of first contact between the droplet and the pool, with $t_i = D/U$ the inertial time. The simulations show that the droplet pushes liquid away on the trailing side (see top panel of figure \ref{fig:timelapse}a), effectively creating an air layer between the pool surface and the impacting droplet. This air layer marks the start of a cavity forming below the pool surface, as can be observed in the side view (see mid panel in figure \ref{fig:timelapse}a). The deepest point of this initial cavity starts on the left side and ramps up along the edge of the impacting droplet towards the pool surface, see mid and top view. Unfortunately, at this stage the cavity is too small to be captured in the experiments as can be seen in the bottom panel. In the simulations, the impact causes the rise of the ejecta sheet on the leading side (see mid panel in figure \ref{fig:timelapse}a). A similar ejecta sheet is observed in the experiment (bottom panel), but the resolution is too low to clearly distinguish the ejecta sheet from the droplet. 

In figure \ref{fig:timelapse}b, the ejecta sheet rises further above the surface and fingers appear on the peripheral side of the rim in the simulation. The droplet is now fully submerged in the pool and the cavity has grown to cover the complete impact area, as can be seen from both the side-view and the top-view. In the simulation, the fingers on the ejecta sheet break up and result in satellite droplets, which is clearly visible in the top-view (top-panel). This break up is not clearly visible in the experiment (bottom-panel), but the overall shape of the ejecta sheet is similar to the simulation. 
\begin{figure*}
\centering
\includegraphics[width=0.99\textwidth]{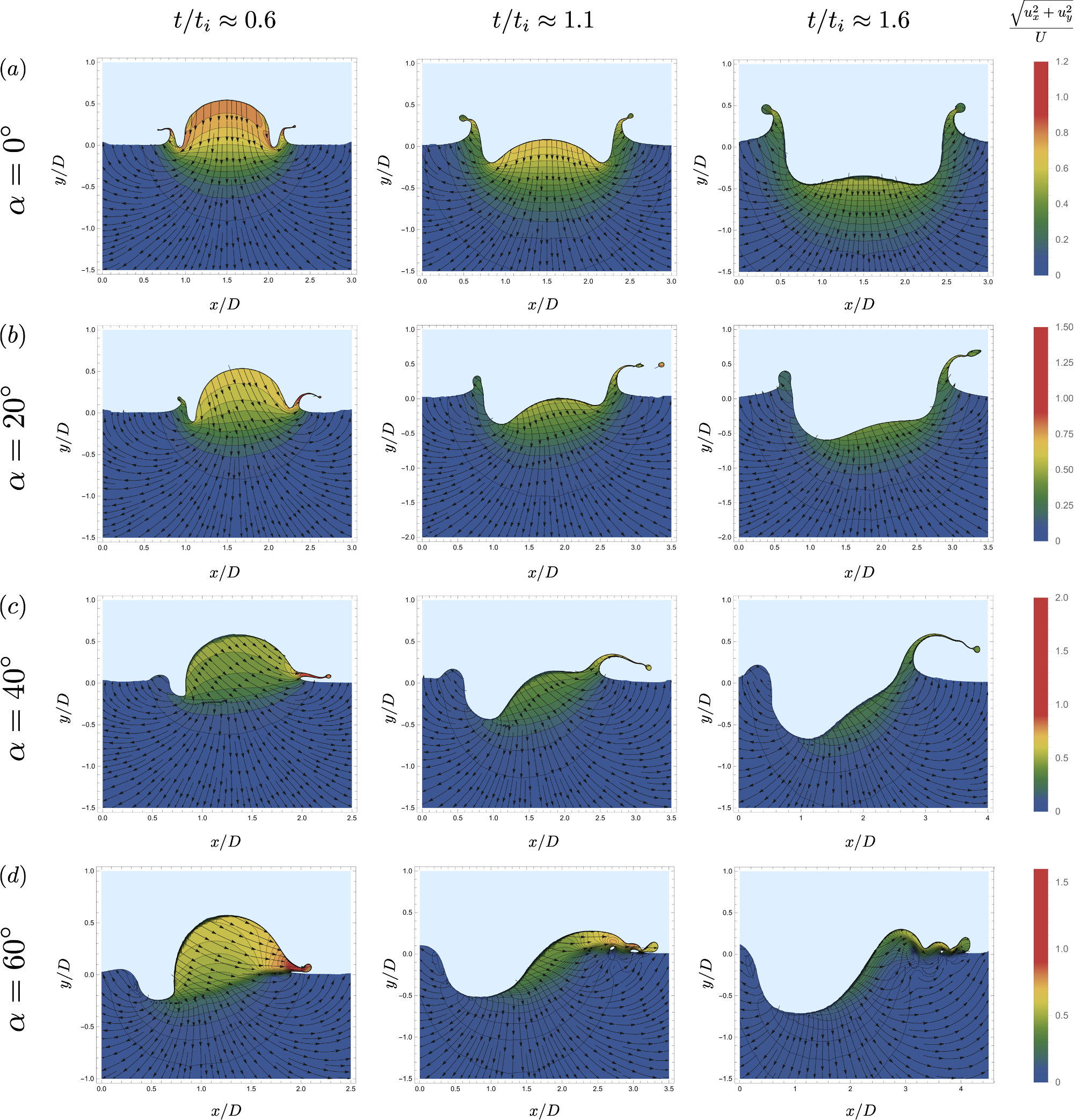}
\caption{\label{fig:vectorplot} A cross-section of the velocity magnitude and streamlines in the liquid phase at the mirror plane ($z=0$) (see figure \ref{fig:sketch}) for $\text{We}=400$ at different angles $\alpha$ and different times $t/t_i$. The line $y=0$ denotes the base level of the pool. The velocity magnitudes (color bar) are scaled by the impact velocity $U$ (note that color bar changes for each row). The light blue color indicates the gas phase.}
\end{figure*}
\begin{figure*}
\centering
\includegraphics[width=0.99\textwidth]{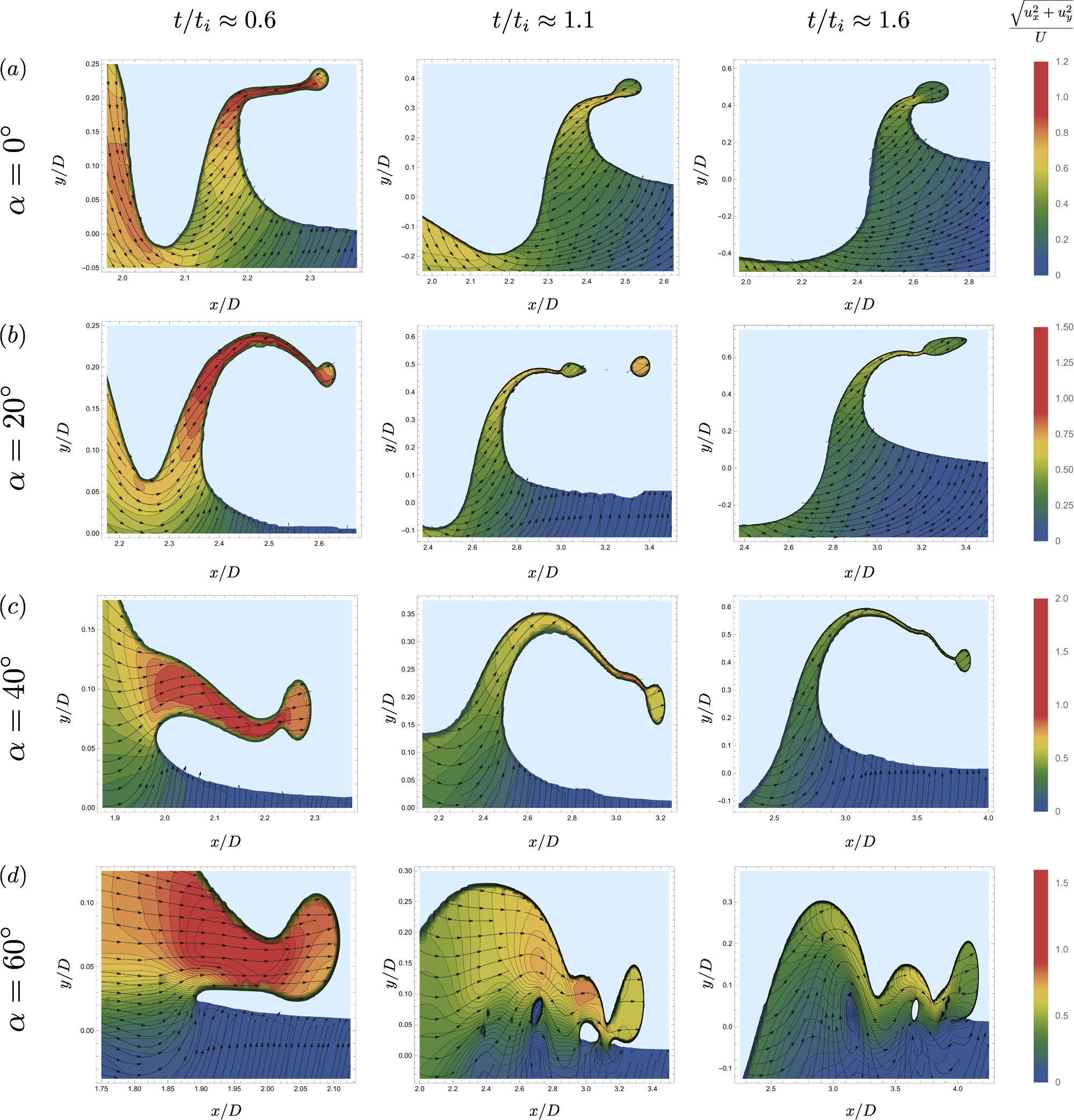}
\caption{\label{fig:vectorplotzoom} A zoom of the cross-section of the velocity magnitude and streamlines of the ejecta sheet on the leading side of the impacting droplet at the mirror plane ($z=0$) (see figure \ref{fig:sketch}) for $\text{We}=400$ at different angles $\alpha$ and different times $t/t_i$. The line $y=0$ denotes the base level of the pool. The velocity magnitudes (color bar) are scaled by the impact velocity $U$ (note that color bar changes for each row). The light blue color indicates the gas phase.}
\end{figure*}
Figure \ref{fig:timelapse}c shows the start of a capillary wave on the trailing side of the cavity that gradually deforms the cavity into a hemispherical shape at later times, see figure \ref{fig:timelapse}d and figure \ref{fig:timelapse}e. Above the pool, the satellite droplets have clearly separated from the ejecta sheet in both the simulation and experiment, which is now best described as a single-sided crown. The fingers are retracting back to the rim of the crown. As the crown itself is also retracting, this causes the fingers to move towards the leading side of the crown, which is clearly visible in the top view panels.

In figure \ref{fig:timelapse}d the retraction of the crown continues while there is only one large finger remaining on the leading side of the crown, which is observed in both numerics and experiments. Finally, in figure \ref{fig:timelapse}e the cavity has reached its maximum expansion. As the remaining finger collapses into the cavity, a capillary wave is triggered. At later times (not shown here) this wave causes fluctuations in the cavity shape.
\subsection{\label{sec:dynamics}Impact dynamics}
In the simulation we have access to the pressure and velocity fields induced in the liquid, which gives a detailed view of the early impact dynamics. Figure \ref{fig:vectorplot} shows a cross-section of the velocity magnitude and streamlines during impact at the mirror plane ($z=0$) for different times $t/t_i$ and angles $\alpha$ at constant $We=400$. In figure \ref{fig:vectorplotzoom} we zoom in on the ejecta sheet region on the leading side of the impacting droplet.

Figure \ref{fig:vectorplot}a shows a perpendicular impact event and serves as a reference case for all consecutive oblique impacts. In the first figure of this row, the droplet has not yet fully submerged into the pool and most of the momentum is still concentrated in the impacting droplet. Inside the pool the streamlines show a symmetric velocity field around the point of impact and bend towards the pool surface. We note that there is a small symmetry deviation between the leading and trailing side of the impacting droplet, which is the result of the adaptive grid refinement strategy used in the simulations. An ejecta sheet is formed on both sides of the impacting droplet, which is more clearly visible in figure \ref{fig:vectorplotzoom}a. The maximum velocity is reached where the ejecta sheet is thinnest. At the edge of the ejecta sheet a cylindrical rim forms due to surface tension forces. At later stages, the droplet fully submerges into the pool and a symmetrically shaped cavity forms. The ejecta sheet keeps rising until it retracts at later stages (not shown).

Figure \ref{fig:vectorplot}b shows an oblique impact event at $\alpha = 20^\circ$. The ejecta sheet is no longer symmetric but larger on the leading-side than on the trailing side. Figure \ref{fig:vectorplotzoom}b shows that the initial velocity magnitude inside the ejecta sheet on the leading side is about $50$\% higher than the original impact speed of the droplet. Furthermore, the maximum velocity in the ejecta sheet exceeds the one achieved during perpendicular impact, see top row. The second figure in this row shows that a satellite droplet is created and as a result the tip now has an oval shape, also seen in the zoomed figure \ref{fig:vectorplotzoom}b. Below the pool surface a cavity starts to form, which grows fastest in the direction of the impacting droplet as shown by the streamlines and velocity magnitudes. Interestingly, the far-field velocity streamlines still resemble the far-field streamlines of the perpendicular case.

The dynamics of an impacting droplet at $\alpha=40^\circ$ is shown in figure \ref{fig:vectorplot}c. During impact a small air layer is entrapped between the pool and droplet, which leads to a set of bubbles concentrated on the left side along the interface of the coalescing droplet with the pool. These bubbles have a local effect on the velocity field which is clearly visible in the first figure of \ref{fig:vectorplot}c. The ejecta sheet on the trailing side is now better described as a liquid bump and carries little momentum. Instead, most of the momentum is concentrated in the ejecta sheet on the leading side where the velocity magnitude is twice of the original impact speed of the droplet. In contrast to the perpendicular case, the ejecta sheet initially grows in the direction along the pool surface but later on deflects upwards, see figure \ref{fig:vectorplotzoom}c. We note that the evolution and shape of the ejecta sheet in figure \ref{fig:vectorplotzoom}b and \ref{fig:vectorplotzoom}c may have a grid dependence, i.e. depends on the grid refinement history. Unfortunately, increasing the maximal grid resolution to $r=13$ was not feasible with the computational resources available to us today. The bubbles entrapped in the cavity do not seem to affect the cavity evolution, which similar to previous cases grows faster in the direction of the impacting droplet. Again we note that the far field streamlines resemble those of the perpendicular impact.

Finally, in figure \ref{fig:vectorplot}d we show an impacting droplet at $\alpha=60^\circ$. Similarly to the previous case bubbles appear on the left-hand side of the interface between the impacting droplet and pool. During impact the droplet partly coalesces with the pool but also partly slides along the surface of the pool, as seen in the first figure of \ref{fig:vectorplot}d. The contour lines of the velocity field confirm this observation, since they form a closely packed layer between the impacting droplet and the pool in which the magnitude and direction rapidly change. The ejecta sheet on the leading side is thicker compared to the previous cases and expands parallel to the surface, as clearly visible in figure \ref{fig:vectorplotzoom}d. Here, the ejecta sheet does not change direction over time and remains parallel to the surface in all plotted stages. The tip of the sheet eventually falls down onto the pool (see second figure of \ref{fig:vectorplotzoom}d). The secondary impact of this sheet onto the pool entraps another air layer on the leading side. Furthermore, the coalescence results in a complex wave pattern on the surface as shown by last figure of \ref{fig:vectorplotzoom}d. The cavity does no longer resemble that of the perpendicular case and even the far-field streamlines (on this scale) start to show asymmetry.
\begin{figure}
 \centering
\includegraphics[width=0.99\textwidth]{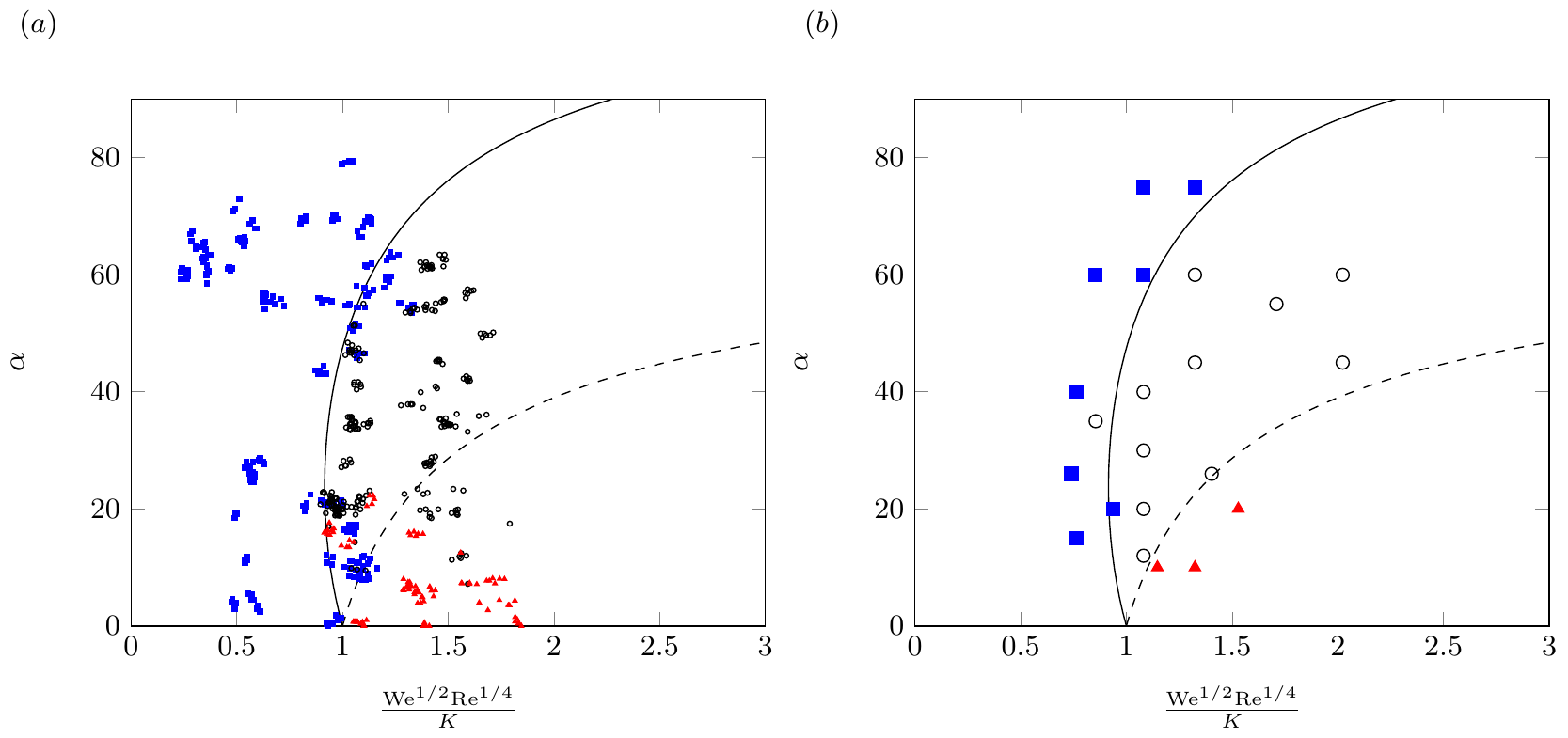}
  \caption{\label{fig:phasediagram}Phase diagram of the impact
behaviour as a function of the dimensionless splashing parameter $\text{We}^{1/2}\text{Re}^{1/4}/K$ and impact angle $\alpha$ for both the experimental results by \citet{PhysRevFluids.2.083602} (a) and the simulation results (b), where $K$ is the critical splashing number. The blue filled squares, black open circles and red closed triangles depict deposition, single-side splashing and omni-directional splashing respectively. The solid and dotted lines are a scaling model that predict the transitions from deposition to single-sided splashing and from single-sided splashing to omni-directional splashing, see main text. In both panels we used $c=0.44$ for the fit parameter measuring the mass-flow distribution around the crown.}
\end{figure}
\subsection{\label{sec:splashingregimes}Splashing threshold}
The three impact phenomena (deposition, single-sided splashing, omni-directional splashing) observed can be quantified in a phase diagram. Figure \ref{fig:phasediagram} shows such a splashing phase diagram in terms of the impact parameters $\text{We}^{1/2}\text{Re}^{1/4}/K$ and $\alpha$ (as discussed by \citet{PhysRevFluids.2.083602}), where $K$ is the critical splashing number. Figure \ref{fig:phasediagram}a shows the experimental results by \citet{PhysRevFluids.2.083602} and figure \ref{fig:phasediagram}b shows the simulation results. The transitions between these different impact phenomena (denoted by the black solid and dotted lines) can be explained by the scaling argument derived by \citet{PhysRevFluids.2.083602}. For completeness, we briefly repeat the argument here.

We assume that the mass flow into the crown is directly proportional to the mass flow into the pool \citep{doi:10.1063/1.1572815}. Since the droplet impacts under an angle $\alpha$, the leading side of the crown accumulates more mass than the trailing side. The mass balance of the crown reads \citep{PhysRevFluids.2.083602}
\begin{equation}
\rho_l D^2 U_\bot \pm \rho_l c D^2 U_\parallel \sim \rho_l e D V,
\end{equation}
where the left-hand side represents the mass flow into the leading ($+$) and trailing ($-$) side of the crown, $e \ll D$ is the thickness of the crown at the pool surface, $V$ is the ejection velocity of the crown and $c$ is a fit parameter that accounts for the exact mass-flow distribution over the crown. To define the splashing threshold it is assumed that breakup of the crown occurs when the crown ejection velocity $V$, scaled by the Taylor-Culick velocity $V_{\text{TC}} \sim \sqrt{\frac{\sigma}{\rho_l e}}$, exceeds a critical value of the splashing number $K$ \citep{doi:10.1063/1.1572815, PhysRevFluids.2.083602}. By using $e \sim \sqrt{\nu D / U}$ for the thickness of the crown at its base \citep{thoroddsen_2002}, we find as splashing criterion of the crown \citep{PhysRevFluids.2.083602}
\begin{equation}
\frac{\text{We}^{1/2} \text{Re}^{1/4}}{K}\cos(\alpha)^{5/4}\bigg[1\pm c\tan(\alpha)\bigg] > 1.
\label{eq:splashingcriterion}
\end{equation}
\begin{figure} 
\centering
\includegraphics[width=0.5\textwidth]{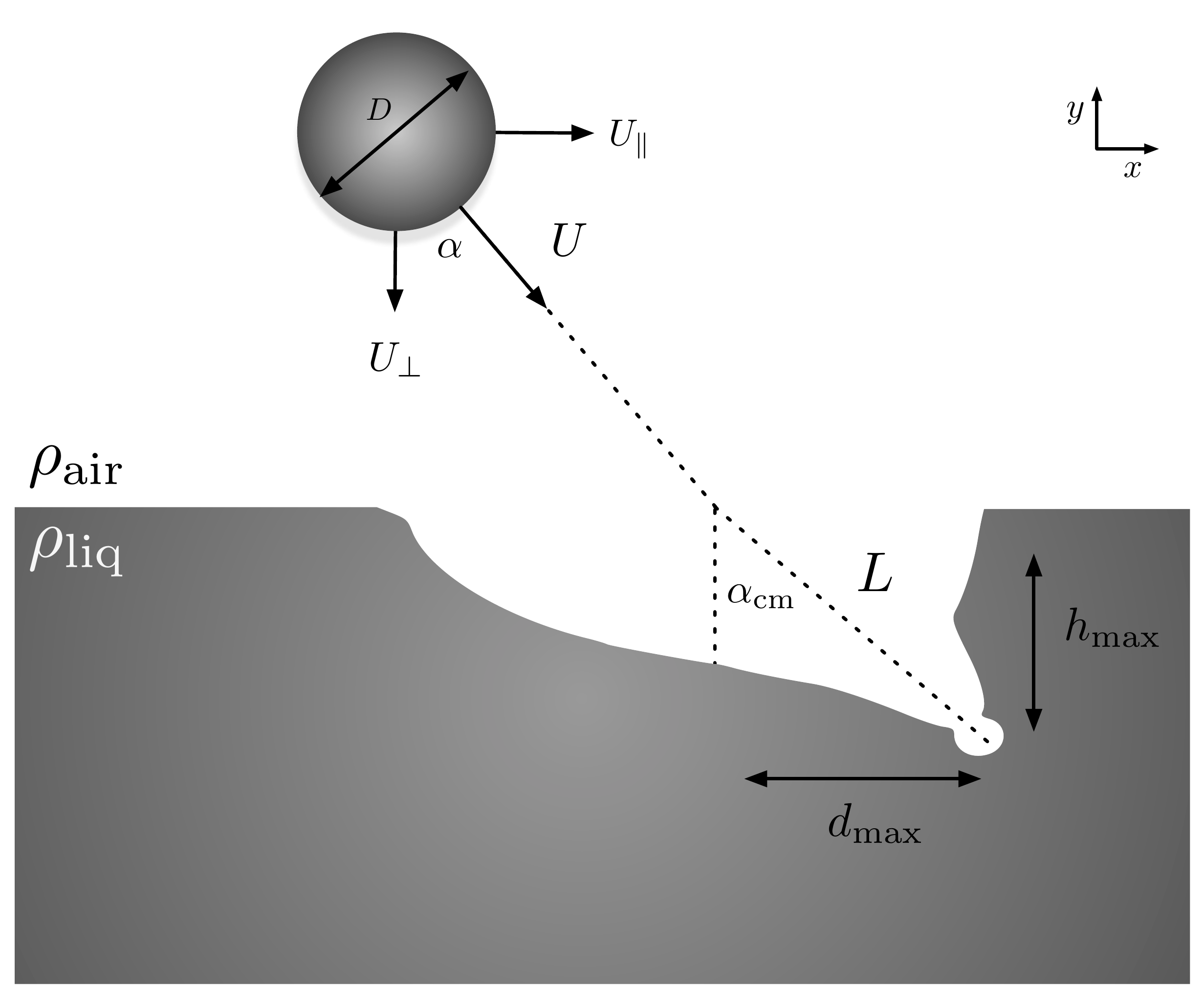}
\caption{\label{fig:sketchdimensions} Schematic side-view of the moment when the cavity reaches its maximum depth $d_\text{max}$, height $h_\text{max}$ and angle $\alpha_\text{cm}$.}
\end{figure} 

\begin{figure} 
\centering
\includegraphics[width=0.7\textwidth]{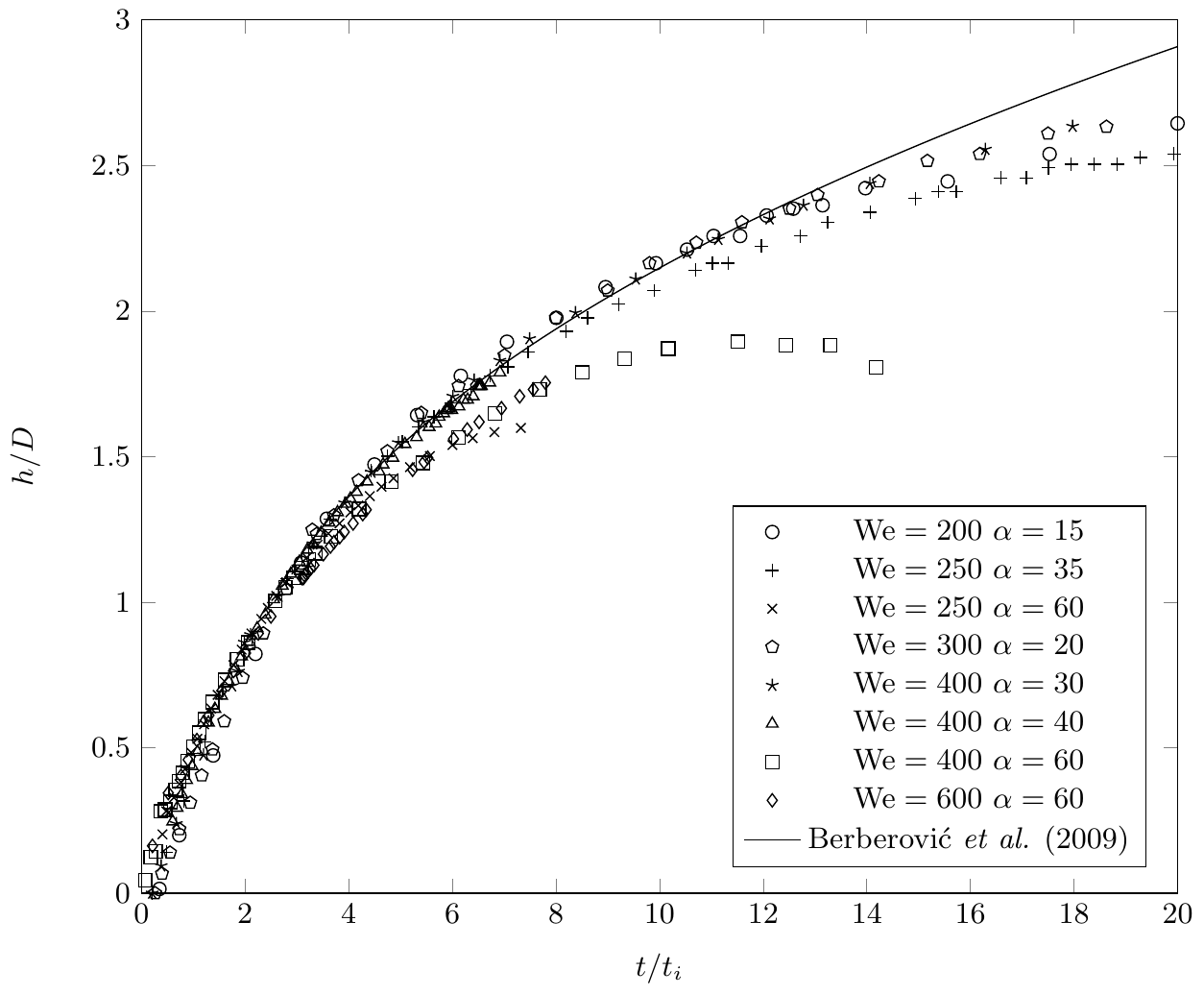}
\caption{\label{fig:cavitydynamics} Cavity depth $h/D$ as function of time $t/t_i$ for different Weber numbers and impact angles (symbols). The solid line corresponds to the theoretical model (\ref{eq:cavityevolution}) as derived by \citet{PhysRevE.79.036306}. }
\end{figure}
\begin{figure} 
\centering
\includegraphics[width=0.99\textwidth]{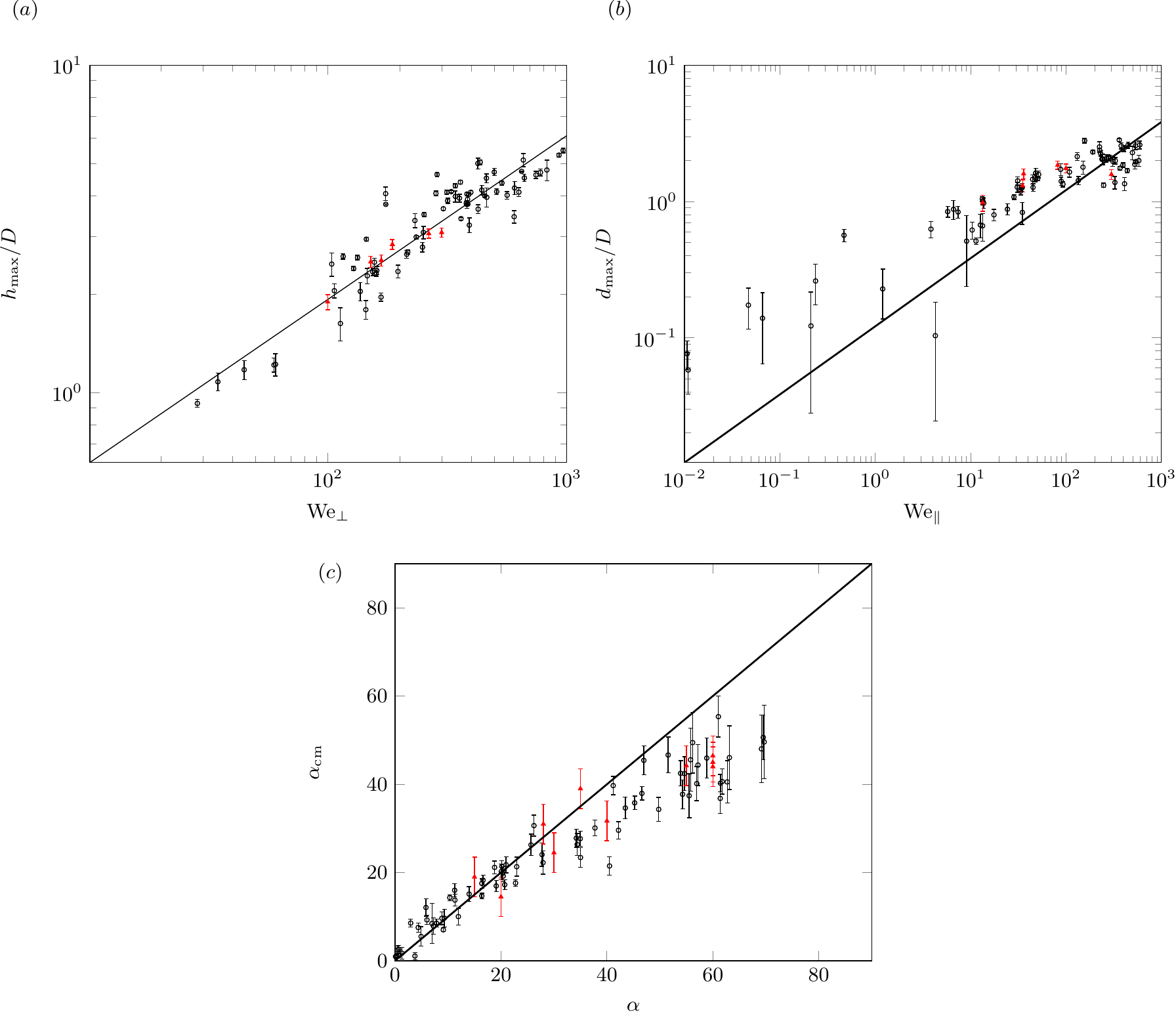}
\caption{\label{fig:dimensions}(a) Double-logarithmic plot of the maximum cavity depth $h_{\text{max}}$ as function of the Weber number $\text{We}_{\bot}$ and (b) the maximum displacement $d_\text{max}$ as function of the parallel Weber number $\text{We}_{\parallel}$. The solid line in (a) and (b) denotes the scaling law (\ref{eq:energyscaleh},\ref{eq:energyscaled}) with prefactors of $0.19$ and $0.12$ taken from experiment and numerical results. (c) Maximum cavity angle $\alpha_{\text{cm}}$ as function of the impact angle $\alpha$. The solid line in (c) is the line with $\alpha_\text{cm} = \alpha$. The open black circles correspond to the experiments and the closed red triangles to the simulations. The error bars in the simulation data are a result of the uncertainty in the cavity depth and displacement, which are subject to fluctuations.}
\end{figure}
The value of $K$ is determined from a perpendicular impact event ($\alpha=0$) and used for all other impacts. In the simulation this leads to $K\approx 104$ and $K\approx 130$ for the experiment \citep{PhysRevFluids.2.083602}. We note that the critical splashing number is dependent on the spatial and temporal resolution of the simulation, i.e. sensitive to under resolution, but has converged at the grid resolution used as described in section \ref{sec:numericalmethod}. The fitting parameter c is set to $c=0.44$, following \citet{PhysRevFluids.2.083602}.

The splashing criterion (\ref{eq:splashingcriterion}) predicts two transitions in the phase space: (i) a transition from deposition to single-sided splashing when the splashing criterion is met on a single side of the crown (the solid line in figure \ref{fig:phasediagram}) and (ii) a transition from single-sided splashing to omni-directional splashing when the criterion is fulfilled on both sides of the crown (the dashed line in figure \ref{fig:phasediagram}). We note that (\ref{eq:splashingcriterion}) is a geometric argument and merely gives a dimensionless scale for which we can expect breakup to happen based on the splashing criterion in the perpendicular case.
%\textcolor{red}{The breakup dynamics of the crown for different Weber numbers, Reynolds numbers and impact angles is extremely nonlinear will be discussed in \ref{sec:cavitydimensions}.} 

We find good quantitative agreement between the simulations, the experiments and the scaling model in all splashing regimes. The experiments show a zone where all three impact behaviours overlap for $\frac{\text{We}^{1/2}\text{Re}^{1/4}}{K} \approx 1$ and $\alpha < 20^\circ$ (see panel a) \citep{PhysRevFluids.2.083602} which is absent in both the model and the simulations (panel b). This impact region is sensitive to small variations in the impact parameters in the experiment and the two-dimensional view makes it difficult to visually distinguish between different impact behaviours. The simulation provides full 3D insight in the impact dynamics and therefore allows for a better judgment on the splashing behaviour in this regime.

\subsection{\label{sec:cavityevolution}Cavity dynamics}
We now turn to a quantitative analysis of the cavity dynamics. A schematic view of the cavity shape is given in figure \ref{fig:sketchdimensions}. We define the maximum cavity dimensions, similar to \citet{PhysRevFluids.2.083602}, by the maximum cavity depth $h_{\text{max}}$, maximum cavity displacement $d_{\text{max}}$ and the maximum cavity angle $\alpha_{cm}$. 

The cavity depth $h$ as function of time $t/t_i$ is plotted in figure \ref{fig:cavitydynamics} for different Weber numbers and impact angles. It turns out that the cavity dynamics at early times ($t/t_i < 4$) follows the same trend for all impact angles studied. This observation inspired us to describe the cavity depth evolution by a model that was previously derived for perpendicular impacts \citep{PhysRevE.82.036319,PhysRevE.79.036306}. Here, the cavity depth is modeled as a hemisphere where the unsteady Bernoulli equation together with a balance of stresses at the cavity interface is used to describe its evolution. In the limiting case of $\text{We}\gg 1$ and $\text{Fr}\gg 1$ the evolution of the cavity depth reads \citep{PhysRevE.82.036319,PhysRevE.79.036306}
\begin{align}
\frac{\mathrm{d}^2 h}{\mathrm{d} t^2} = -\frac{3}{2 h}\left(\frac{\mathrm{d} h}{\mathrm{d} t}\right)^2,
\end{align}
which has the solution
\begin{align}
h(t/t_i)/D= c_1(5 t/t_i - c_2)^{2/5},
\label{eq:cavityevolution}
\end{align}
where $c_1=0.47$ and $c_2=5.87$ are constants determined by a least-square fit to the data with impact angles $\leq 30^\circ$ (i.e. nearly perpendicular) for $2<t/t_i<15$, which is the regime where the model applies. The model captures the simulation results with good accuracy for impact angles smaller than $30^\circ$. When the impact angle is larger than $30^\circ$ the cavity can no longer be modeled as an expanding hemisphere. Instead, due to the large impact angle the cavity is stretched in the direction parallel to the pool surface, leading to a cone-shaped cavity. Indeed, in figure \ref{fig:vectorplot} we observe that for $40^\circ$ the streamlines start to deviate from the perpendicular impact case. As a result the cavity depth becomes smaller than the depth predicted by the hemispherical model, as is clear from figure \ref{fig:cavitydynamics}.
\subsection{\label{sec:cavitydimensions}Cavity dimensions} 
Figure \ref{fig:dimensions} gives an overview of the cavity dimensions $h_{\text{max}}$ and $d_\text{max}$ as function of $\text{We}_\parallel$, $\text{We}_\perp$ and $\alpha$. In the top two panels we show a double-logarithmic plot of the maximum cavity depth $h_{\text{max}}$ (left) and the maximum cavity displacement $d_\text{max}$ (right) as function of respectively the perpendicular and parallel Weber number. The black open circles represent the experimental results by \citet{PhysRevFluids.2.083602} and the red closed triangles are the numerical results. The solid lines corresponds to a scaling argument that relates the kinetic energy of the droplet to the surface energy the cavity acquires, which leads to \citep{PhysRevFluids.2.083602}
\begin{align}
\frac{h_\text{max}}{D} \sim \text{We}^{1/2}_\parallel \label{eq:energyscaleh}, \\
\frac{d_\text{max}}{D} \sim \text{We}^{1/2}_\perp, \label{eq:energyscaled}
\end{align}
where it was assumed that $\alpha_{\text{cm}} \approx \alpha$. The simulation results are in agreement with the experimental data. The scaling argument confirms the trend for both the cavity depth and cavity displacement. There is however a large uncertainty for low Weber numbers in the experimental data for the cavity displacement, since in this regime the displacement is hard to measure \cite{PhysRevFluids.2.083602}.

Figure \ref{fig:dimensions}c shows a double-logarithmic plot of the maximum cavity angle $\alpha_\text{cm}$ as function of the impact angle $\alpha$. Similar to the experimental results, we observe that the maximum cavity angle increases linearly with increasing impact angle up to $\alpha \approx 30^\circ$.  For larger impact angles, the cavity angle is smaller than the impact angle. This effect was also observed in the experiments by \citet{PhysRevFluids.2.083602}, and was attributed to energy dissipation by waves tangential to the surface. 
\section{\label{sec:conclusion}Discussion \& Conclusion}
We presented a numerical study of oblique droplet impact onto a deep liquid pool by using the adaptive volume-of-fluid solver Basilisk. The splashing behaviour and cavity dynamics after impact were quantified as function the Weber number and the impact angle. The numerical results were compared to recent experimental work by \citet{PhysRevFluids.2.083602}. We found good qualitative and quantitative agreement between the numerical and experimental results on the dynamics of both the sheet above and the cavity below the pool surface for a broad range of Weber numbers and impact angles. In particular, the splashing threshold, the maximum cavity depth, maximum cavity displacement and cavity angle were found to be in good agreement with the experiments.

The spatial and temporal resolution of the simulations was chosen such that the sheet dynamics up to the moment of breakup and the cavity dynamics up to the moment the maximum cavity depth is reached are converged. These resolution parameters are bound by the time and computational resources available, which limited us to a maximum Weber number $\mathcal{O}(10^3)$. In the higher Weber number regime sharp interfaces and large velocity gradients dominate the dynamics. Insufficient resolution may therefore cause the sheet and cavity dynamics to be captured incorrectly.

To be able to increase the grid resolution while maintaining a reasonable computation time we imposed a symmetry condition in the $y-z$ plane (see \ref{fig:sketch}). We found that this symmetry condition improved the convergence of the overall sheet dynamics. This constraint however may restrict the Rayleigh-Taylor instability from developing correctly at the rim of the crown. For a detailed analysis of the fingering instability the constraint should therefore be released.

The numerical simulations provide a valuable addition to the experimental results, because they capture a fully 3D view of the impact phenomenon which allows for a better judgment on the single-sided and omni-directional splashing thresholds. In addition, the method provides access to velocity and pressure fields which are not available in the experiments and allow to obtain more insight into the cavity and ejecta-sheet dynamics. Moreover, the method can be used to explore impact parameters that are hard to reach experimentally, e.g. very large impact angles or high droplet velocities.

\begin{acknowledgments}
This work is part of an Industrial Partnership Programme of the Netherlands Organization for Scientific Research (NWO). This research programme is co-financed by ASML. We also would like to thank Marise Gielen, Alexander Klein and Wim-Paul Breugem for feedback on this work.
\end{acknowledgments}

\bibliography{references}

\end{document}